\documentclass[letterpaper,twocolumn,american,showpacs,pra,aps,superscriptaddress,eqsecnum]{revtex4}
\usepackage[T1]{fontenc}
\usepackage[latin1]{inputenc}
\usepackage{bm}
\usepackage{amsmath}
\usepackage{babel}
\usepackage{graphicx}
\usepackage{amssymb}
\makeatletter
\usepackage{pifont}
\makeatother

\begin{document}

\title{State space structure and entanglement of rotationally invariant spin systems}

\author{Heinz-Peter Breuer}

\email{breuer@physik.uni-freiburg.de}

\affiliation{Physikalisches Institut, Universit\"at Freiburg,
             Hermann-Herder-Str.~3, D-79104 Freiburg, Germany}

\date{\today}

\begin{abstract}
We investigate the structure of $\mathrm{SO(3)}$-invariant quantum
systems which are composed of two particles with spins $j_1$ and
$j_2$. The states of the composite spin system are represented by
means of two complete sets of rotationally invariant operators,
namely by the projections $P_J$ onto the eigenspaces of the total
angular momentum $J$, and by certain invariant operators $Q_K$
which are built out of spherical tensor operators of rank $K$. It
is shown that these representations are connected by an orthogonal
matrix whose elements are expressible in terms of Wigner's $6$-$j$
symbols. The operation of the partial time reversal of the
combined spin system is demonstrated to be diagonal in the
$Q_K$-representation. These results are employed to obtain a
complete characterization of spin systems with $j_1=1$ and
arbitrary $j_2 \geq 1$. We prove that the Peres-Horodecki
criterion of positive partial transposition (PPT) is necessary and
sufficient for separability if $j_2$ is an integer, while for
half-integer spins $j_2$ there always exist entangled PPT states
(bound entanglement). We construct an optimal entanglement witness
for the case of half-integer spins and design a protocol for the
detection of entangled PPT states through measurements of the
total angular momentum.
\end{abstract}


\maketitle

\section{Introduction}
Entanglement is a basic feature of composite quantum systems
connected to the tensor product structure of the underlying
Hilbert space of states. A mixed state of a bipartite quantum
system described by some density matrix $\rho$ is said to be
entangled or inseparable if $\rho$ cannot be written as a convex
linear combination of product states. Otherwise it is called
classically correlated or separable \cite{WERNER}. The properties
of entangled states are responsible for many of the fascinating
and curious aspects of the quantum world and lie at the core of
many proposed applications in quantum information processing
\cite{ECKERT,ALBER,NIELSEN}.

The general characterization and quantification of entanglement in
mixed quantum states is a highly non-trivial problem. It is even
very difficult in general to formulate simple operational criteria
which allow a unique identification of all separable states of a
given composite system. There do exist, however, many necessary
separability criteria
\cite{PERES,HORODECKI96a,HORODECKI99,CERF,KEMPE,RUDOLPH03,CHEN,TERHAL,LEWENSTEIN00}.
A simple and, in fact, very strong criterion is the
Peres-Horodecki criterion \cite{PERES,HORODECKI96a} which states
that a necessary condition for a given density matrix $\rho$ to be
separable is that it has a positive partial transposition (PPT
states). It is known that this criterion is necessary and
sufficient for certain low-dimensional systems, while it is only
necessary in higher dimensions \cite{HORODECKI96a}.

The analysis of the entanglement structure is greatly facilitated
through the introduction of symmetries, i.~e., if one restricts to
those states of the composite system which are invariant under
certain groups of symmetry transformations. Important examples in
this context are the manifolds of the Werner states \cite{WERNER},
of the isotropic states \cite{RAINS,HORODECKI99} and of the
orthogonal states \cite{VOLLBRECHT}. Here, we investigate
entanglement under the symmetry group $\mathrm{SO(3)}$ of proper
three-dimensional rotations of the coordinate axes. More
precisely, we consider the problem of mixed state entanglement of
systems which are composed of two particles with spins $j_1$ and
$j_2$, and which are invariant under product representations of
the group $\mathrm{SO(3)}$ or, equivalently, of the covering group
$\mathrm{SU(2)}$. A basic tool of our analysis is the work of
Vollbrecht and Werner \cite{VOLLBRECHT} which provides a general
scheme for the treatment of entanglement under given symmetry
groups.

Mixed $\mathrm{SO(3)}$-invariant states of composite systems
arise, for example, from the interaction of open systems
\cite{TheWork} with isotropic environments \cite{GORINI}. Their
analysis is of great importance and leads to many applications. As
examples we mention investigations on the connection between
quantum phase transitions and the behaviour of entanglement
measures (see, e.~g., \cite{OSTERLOH,OSBORNE}), the analysis
of entanglement of $\mathrm{SU(2)}$-invariant multiphoton states
generated by parametric down-conversion \cite{DURKIN}, and studies 
of the entanglement of formation \cite{CAVES}. The
technique of this paper could also be relevant for the
characterization of quantum correlations in Fermionic or Bosonic
systems developed recently \cite{SCHLIEMANN01,BRUSS}.

The Hilbert space of a system which is composed of two particles
with spins $j_1$ and $j_2$ is given by the tensor product
${\mathbb C}^{N_1}\otimes{\mathbb C}^{N_2}$, where $N_1=2j_1+1$
and $N_2=2j_2+1$ are the dimensions of the local spin spaces. We
call such a system an $N_1 \otimes N_2$ system. Throughout the
paper we will assume that $j_1 \leq j_2$, i.~e., $N_1 \leq N_2$.

According to the Peres-Horodecki criterion
\cite{PERES,HORODECKI96a} the cases of $2\otimes 2$ and $2\otimes
3$ systems are trivial: It is known that in these cases the PPT
criterion is necessary and sufficient for all states, i.~e., even
for states which are not invariant under rotations. Schliemann
\cite{SCHLIEMANN1} has shown recently that the PPT criterion is
also necessary and sufficient for $\mathrm{SO(3)}$-invariant
$2\otimes N_2$ systems with arbitrary $N_2$. The case of $3
\otimes 3$ systems has been treated by Vollbrecht and Werner
\cite{VOLLBRECHT}, who proved that the PPT criterion is again
necessary and sufficient for separability. For $4\otimes 4$
systems a qualitatively new situation arises: It has been
demonstrated in \cite{NtensorN} that the PPT criterion is not
sufficient and that the entangled PPT states form a
three-dimensional manifold which is isomorphic to a prism. In the
present work we investigate the important special case of $3
\otimes N_2$ systems with arbitrary $N_2$.

The method developed in \cite{NtensorN} enables the treatment of
the case of equal spins $j_1=j_2$. In this paper we extend this
method to arbitrary spins $j_1$ and $j_2$. For the analysis of
entanglement under $\mathrm{SO(3)}$-symmetry it is advantageous to
replace the transposition used in the PPT criterion by another
unitarily equivalent operation, namely by the antiunitary
transformation of the time reversal. The reason for this fact is
that the operation of the time reversal of states commutes with
the representations of the rotation group.

There are two natural representations of rotationally invariant
states. The first one uses the fact that any invariant state can
be written as a unique convex linear combination of the
projections $P_J$ onto the eigenspaces of the total angular
momentum $J$ of the composite spin system. The advantage of this
representation is that it leads to very simple conditions
expressing the positivity and the normalization of physical
states. However, the set of the PPT states is most easily
determined in another representation which employs the irreducible
spherical tensor operators of spin-$j$ particles. We will
construct a complete system of invariant operators $Q_K$ which are
built out of the spherical tensors of rank $K$. Any invariant
state of the composite spin system can then be written as a unique
linear combination of the $Q_K$. The introduction of the invariant
operators $Q_K$ generalizes the ideas of Schliemann
\cite{SCHLIEMANN1,SCHLIEMANN2}, who has developed a representation
of $\mathrm{SU(2)}$-invariant states by means of spin-spin
correlators and has formulated various separability conditions and
sum rules in terms of these correlators.

The paper is organized as follows. The representations of
$\mathrm{SO(3)}$-invariant states in terms of the invariant
operators $P_J$ and $Q_K$ are constructed in Sec.~\ref{DD-INVAR}.
We also derive in this section the linear transformation which
connects these representations and show that it is given by an
orthogonal matrix whose elements are determined by Wigner's
$6$-$j$ symbols. The behaviour of states under partial time
reversal and the construction of the set of the invariant
separable states are discussed in Sec.~\ref{INVAR-SEP}.

The general theory is then applied in Sec.~\ref{3tensorN} to the
case of $3 \otimes N_2$ systems with arbitrary $N_2$. We prove
that the PPT criterion represents a necessary and sufficient
separability condition for $3 \otimes N_2$ systems if and only if
$N_2$ is odd. Thus, for integer spins $j_2$ all PPT states are
separable, while for half-integer spins $j_2$ there always exist
entangled PPT states. This fact has already been conjectured by
Hendriks \cite{HENDRIKS} on the basis of a detailed numerical
investigation. We also show that for half-integer $j_2$ the
boundary of the separability region is curved. Finally,
Sec.~\ref{CONCLU} contains a discussion of the results and some
conclusions. In particular, we construct an optimal entanglement
witness for the case of half-integer spins and exploit this
witness to design a protocol which allows the detection of
entangled PPT states through measurements of the total angular
momentum.

\section{Representations of $\mathrm{SO(3)}$-invariant states}\label{DD-INVAR}
We consider two particles with spins $j_1$ and $j_2$ and
corresponding angular momentum operators $\hat{\bm{j}}^{(1)}$ and
$\hat{\bm{j}}^{(2)}$. The Hilbert space ${\mathbb C}^{N_1}$ of the
first particle is spanned by the common eigenstates
$|j_1,m_1\rangle$ of the square of $\hat{\bm{j}}^{(1)}$ and of
$\hat{\bm{j}}^{(1)}_z$, where $N_1\equiv 2j_1+1$ and
$m_1=-j_1,\ldots,+j_1$. Correspondingly, the Hilbert space
${\mathbb C}^{N_2}$ of the second particle is spanned by the
eigenstates $|j_2,m_2\rangle$, where $N_2\equiv 2j_2+1$ and
$m_2=-j_2,\ldots,+j_2$.

The Hilbert space of the total system composed of both particles
is given by the tensor product ${\mathbb C}^{N_1}\otimes{\mathbb
C}^{N_2}$. The angular momentum operator of the composite system
is defined by:
\begin{equation}
 \hat{\bm{J}} = \hat{\bm{j}}^{(1)} \otimes I + I \otimes
 \hat{\bm{j}}^{(2)},
\end{equation}
where $I$ denotes the unit matrix. A state of the composite system
is described by a density matrix on the product space, i.~e., by a
positive operator $\rho$ on ${\mathbb C}^{N_1}\otimes{\mathbb
C}^{N_2}$ with unit trace: $\rho \geq 0$, ${\mathrm{tr}}\rho=1$.

The irreducible unitary representation of the group
$\mathrm{SO(3)}$ of proper rotations $R$ on the state space of a
particle with spin $j$ will be denoted by $D^{(j)}(R)$. The
transformation of the states of the composite spin system is then
given by the product representation $D^{(j_1)}(R)\otimes
D^{(j_2)}(R)$. A state $\rho$ of the combined system is said to be
rotationally invariant or $\mathrm{SO(3)}$-invariant if the
relation
\[
 \left[ D^{(j_1)}(R) \otimes D^{(j_2)}(R) \right] \rho
 \left[ D^{(j_1)}(R) \otimes D^{(j_2)}(R) \right]^{\dagger}
 = \rho
\]
holds true for all proper rotations $R\in SO(3)$.

We shall use two different representations of rotationally
invariant states. The first one employs the projection operators
\begin{equation} \label{DEF-PJ}
 P_J = \sum_{M=-J}^{+J} |JM\rangle\langle JM|,
\end{equation}
where $|JM\rangle$ denotes the common eigenstate of the square of
the total angular momentum $\hat{\bm{J}}$ and of its $z$-component
$\hat{J}_z$, i.~e., we have
$\hat{\bm{J}}^2|JM\rangle=J(J+1)|JM\rangle$ and
$\hat{J}_z|JM\rangle=M|JM\rangle$. The operator $P_J$ projects
onto the manifold which is spanned by the eigenstates belonging to
a fixed value $J$ of the total angular momentum. According to the
triangular inequality $J$ takes on $N_1$ different values which
may be integer or half-integer valued:
\begin{equation}
 J = j_2-j_1,j_2-j_1+1, \ldots, j_2+j_1.
\end{equation}
It follows from Schur's lemma that any invariant state $\rho$ can
be written as a linear combination of the $P_J$:
\begin{equation} \label{RHO-J}
 \rho = \frac{1}{\sqrt{N_1N_2}} \sum_J \frac{\alpha_J}{\sqrt{2J+1}} P_J.
\end{equation}
Here, the $\alpha_J$ are real parameters and we have introduced
convenient normalization factors of $\sqrt{N_1N_2}$ and
$\sqrt{2J+1}$. In order for Eq.~(\ref{RHO-J}) to represent a true
density matrix the $\alpha_J$ must of course be positive and
normalized appropriately:
\begin{eqnarray}
 \alpha_J &\geq& 0, \label{S-A-1} \\
 {\mathrm{tr}}\rho &=& \sum_J
 \sqrt{\frac{2J+1}{N_1N_2}}\alpha_J = 1. \label{S-A-2}
\end{eqnarray}

Any invariant state $\rho$ is thus uniquely characterized by a
real vector $\bm{\alpha}$ in an $N_1$-dimensional parameter space
${\mathbb R}^{N_1}$ which will be referred to as $\alpha$-space.
The conditions of the positivity and of the normalization of
$\rho$ are expressed by the relations (\ref{S-A-1}) and
(\ref{S-A-2}). We denote the set of all vectors $\bm{\alpha}$
whose components $\alpha_J$ satisfy these relations by
$S^{\alpha}$. Being isomorphic to the set of invariant states,
$S^{\alpha}$ is of course a convex set. We infer from
Eqs.~(\ref{S-A-1}) and (\ref{S-A-2}) that $S^{\alpha}$ represents
an $(N_1-1)$-dimensional simplex.

A useful alternative representation of the invariant states is
obtained by use of a complete system of irreducible spherical
tensor operators (see, e.~g.~\cite{EDMONDS,SCHWINGER}). The tensor
operators which act on the state space of the particle with spin
$j_i$ are written as $T^{(i)}_{K_iq_i}$, where $i=1,2$. The index
$K_i=0,1,\ldots,2j_i$ denotes the rank of the tensor operator. For
a given rank $K_i$ the index $q_i$ takes on the values
$q_i=-K_i,-K_i+1,\ldots,+K_i$. We thus have $(2K_i+1)$ tensor
operators $T^{(i)}_{K_iq_i}$ of rank $K_i$ which transform under
rotations according to an irreducible representation of the
rotation group. The explicit definitions of the tensors and a
brief summary of their properties are given in Appendix
\ref{APP-1}.

Using the tensor operators one defines Hermitian operators $Q_K$
acting on the state space of the composite spin system:
\begin{equation} \label{DEF-QK}
 Q_K = \sum_{q=-K}^{+K} T^{(1)}_{Kq} \otimes T^{(2)\dagger}_{Kq},
\end{equation}
where the index $K$ takes on $N_1$ different integer values:
\begin{equation}
 K = 0,1,\ldots,2j_1.
\end{equation}
It follows from the transformation properties of the tensor
operators that all $Q_K$ are invariant under rotations. For
instance, the operator $Q_0$ is proportional to the identity,
$Q_0=\frac{1}{\sqrt{N_1N_2}} I\otimes I$, while $Q_1$ is
proportional to the invariant scalar product
$\hat{\bm{j}}^{(1)}\cdot \hat{\bm{j}}^{(2)}$ of the spin vectors.

The $Q_K$ defined by Eq.~(\ref{DEF-QK}) form a complete system of
operators. This means that any rotationally invariant Hermitian
operator can be represented as a unique linear combination of the
$Q_K$ in a way analogous to Eq.~(\ref{RHO-J}):
\begin{equation} \label{RHO-K}
 \rho = \frac{1}{\sqrt{N_1N_2}} \sum_K \frac{\beta_K}{\sqrt{2K+1}}
 Q_K.
\end{equation}
Here, we have again introduced appropriate normalization factors
and real parameters $\beta_K$ which form a vector $\bm{\beta}$ in
an $N_1$-dimensional parameter space ${\mathbb R}^{N_1}$ referred
to as $\beta$-space. The operators $Q_K$ satisfy
${\mathrm{tr}}\{Q_KQ_{K'}\}=(2K+1)\delta_{KK'}$. This fact follows
directly from the orthogonality relation (\ref{T-ORTHO}) for the
spherical tensors. The $Q_K$ for $K \neq 0$ are therefore
traceless which leads to the normalization condition
\begin{equation} \label{NORM-BETA}
 {\mathrm{tr}}\rho = \beta_0 = 1.
\end{equation}

The sets $\{P_J\}$ and $\{Q_K\}$ represent complete systems of
invariant operators. The corresponding parameter vectors
$\bm{\alpha}$ and $\bm{\beta}$ must therefore be related by a
linear transformation ${\mathbb R}^{N_1}\mapsto {\mathbb
R}^{N_1}$. We write
\begin{equation}
 {\bm{\beta}} = L {\bm{\alpha}},
\end{equation}
where $L$ is an $(N_1 \times N_1)$ matrix. To find the elements of
this matrix we use Eqs.~(\ref{RHO-J}) and (\ref{RHO-K}) to get
\begin{equation}
 \sum_J \frac{\alpha_J}{\sqrt{2J+1}} P_J
 = \sum_K \frac{\beta_K}{\sqrt{2K+1}} Q_K.
\end{equation}
Multiplying this equation by $Q_{K'}$ and taking the trace we find
that the elements of $L$ are given by
\begin{equation} \label{L-REP}
 L_{KJ} = [(2K+1)(2J+1)]^{-1/2} {\mathrm{tr}}\{Q_KP_J\}.
\end{equation}
This can be expressed as
\begin{equation} \label{L-6j}
 L_{KJ} = \sqrt{(2K+1)(2J+1)} (-1)^{j_1+j_2+J}
 \left\{\begin{array}{ccc}
 j_1 & j_2 & J \\
 j_2 & j_1 & K
 \end{array}\right\}.
\end{equation}
The curly brackets denote a $6$-$j$ symbol introduced by Wigner
\cite{WIGNER} into the quantum theory of angular momentum. A proof
of the relation (\ref{L-6j}) is given in Appendix~\ref{APP-2}. The
$6$-$j$ symbols are scalar quantities which are defined through
invariant sums over products of Clebsch-Gordan coefficients. They
describe the transformation between different coupling schemes for
the addition of three angular momenta \cite{EDMONDS}. Their
properties have been studied in great detail and many closed
formulae, recursion relations and sum rules are known. In
particular, it follows from the sum rules that $L$ represents an
orthogonal $(N_1 \times N_1)$ matrix.

The above results lead to the conclusion that the set of
$\mathrm{SO(3)}$-invariant states is represented in $\beta$-space
by the set
\begin{equation} \label{SA-SB}
 S^{\beta} = L S^{\alpha}.
\end{equation}
The set $S^{\beta}$ is again an $(N_1-1)$-dimensional simplex
which may be constructed by determining the images of the extreme
points of $S^{\alpha}$ under the orthogonal transformation $L$.

The introduction of two parameter spaces is motivated by the
following observations. On the one hand, the set of states is most
easily constructed as a subset in $\alpha$-space. This is due to
the fact that the representation of Eq.~(\ref{RHO-J}) corresponds
to the spectral decomposition of $\rho$ and, therefore, the
requirement of the positivity of $\rho$ immediately leads to the
simple condition (\ref{S-A-1}). On the other hand, the
representation (\ref{RHO-K}) of states in $\beta$-space is much
more suitable for the construction of the set of separable states,
which is due to the fact that the operation of the partial time
reversal is diagonal in the $Q_K$-representation.

\section{Invariant separable states}\label{INVAR-SEP}
A state $\rho$ of the composite spin system is said to be
separable if it is possible to write this state as a convex linear
combination of product states:
\begin{equation} \label{DEF-SEP}
 \rho = \sum_i \lambda_i \rho_i^{(1)} \otimes \rho_i^{(2)},
 \qquad \lambda_i \geq 0, \qquad \sum_i \lambda_i = 1,
\end{equation}
where the $\rho_i^{(1)}$ and $\rho_i^{(2)}$ are normalized states
of the first and of the second spin, respectively \cite{WERNER}.
It is clear that the set in $\beta$-space which represents the
invariant and separable states is a convex subset of $S^{\beta}$.
This subset will be denoted by $S^{\beta}_{\mathrm{sep}}$.

Following the work of Vollbrecht and Werner \cite{VOLLBRECHT} we
define a projection super-operator ($\mathrm{SO(3)}$ twirling) by
means of
\begin{equation} \label{PI-DEF-1}
 \Pi\rho = \int dR \; U(R) \rho U(R)^{\dagger},
\end{equation}
where $U(R)\equiv D^{(j_1)}(R) \otimes D^{(j_2)}(R)$ and the
integration is extended over all group elements $R\in SO(3)$. The
twirl operation maps any state $\rho$ of the composite spin system
to an $\mathrm{SO(3)}$-invariant state $\Pi\rho$. Moreover, if
$\rho$ is separable then also $\Pi\rho$ is separable. In terms of
the invariant operators $P_J$ or $Q_K$ the action of the twirl
operation may be expressed by
\begin{equation} \label{PI-DEF-2}
 \Pi\rho = \sum_J \frac{{\mathrm{tr}}\{P_J\rho\}}{2J+1}P_J
 = \sum_K \frac{{\mathrm{tr}}\{Q_K\rho\}}{2K+1}Q_K.
\end{equation}
It is known that any invariant separable state is a convex linear
combination of $\Pi$-projections of pure product states. Given a
pure product state
\begin{equation} \label{PROD-STATE}
 \rho =
 |\varphi^{(1)}\varphi^{(2)}\rangle\langle\varphi^{(1)}\varphi^{(2)}|,
\end{equation}
Eq.~(\ref{PI-DEF-2}) shows that the corresponding parameters
$\alpha_J$ and $\beta_K$ of its projection $\Pi\rho$ are given by
\begin{eqnarray}
 \alpha_J &=& \sqrt{\frac{N_1N_2}{2J+1}}
 \langle\varphi^{(1)}\varphi^{(2)}|P_J|\varphi^{(1)}\varphi^{(2)}\rangle,
 \label{ALPHA-SEP} \\
 \beta_K &=& \sqrt{\frac{N_1N_2}{2K+1}}
 \langle\varphi^{(1)}\varphi^{(2)}|Q_K|\varphi^{(1)}\varphi^{(2)}\rangle.
 \label{BETA-SEP}
\end{eqnarray}

We introduce into Eq.~(\ref{BETA-SEP}) the definition
(\ref{DEF-QK}) of the $Q_K$ and define the functions
\begin{eqnarray} \label{DEF-FUNC-BETA-K}
 \lefteqn{ \tilde{\beta}_K[\varphi^{(1)},\varphi^{(2)}] } \\
 && = \sqrt{\frac{N_1N_2}{2K+1}} \sum_{q=-K}^{+K}
 \langle\varphi^{(1)}|T^{(1)}_{Kq}|\varphi^{(1)}\rangle
 \langle\varphi^{(2)}|T^{(2)\dagger}_{Kq}|\varphi^{(2)}\rangle.
 \nonumber
\end{eqnarray}
Let us further define $W^{\beta}$ as the range of the parameter
vector $\bm{\beta}$ whose components are given by these functions,
where $|\varphi^{(1)}\rangle \in {\mathbb C}^{N_1}$ and
$|\varphi^{(2)}\rangle\in {\mathbb C}^{N_2}$ run independently
over all normalized states of the first and of the second spin,
respectively:
\begin{equation} \label{DEF-W-BETA}
 W^{\beta} = \left\{ \bm{\beta} \left| \;
 \beta_K=\tilde{\beta}_K[\varphi^{(1)},\varphi^{(2)}], \;
 ||\varphi^{(1,2)}|| = 1 \right. \right\}.
\end{equation}
The set of separable states is then equal to the convex hull of
$W^{\beta}$:
\begin{equation} \label{HULL}
 S^{\beta}_{\mathrm{sep}} = {\mbox{hull}}\left(W^{\beta}\right).
\end{equation}
This means that $S^{\beta}_{\mathrm{sep}}$ is equal to the
smallest convex set which contains $W^{\beta}$.

Within this formulation the problem of constructing
$S^{\beta}_{\mathrm{sep}}$ reduces to the determination of the
convex hull of the range of the functions $\tilde{\beta}_K$. Even
for the present case of a highly symmetric state space this is, in
general, an extremely difficult task. A strong necessary condition
for separability is the Peres-Horodecki criterion
\cite{PERES,HORODECKI96a}. According to this criterion a necessary
condition for a given state $\rho$ to be separable is that its
partial transposition is a positive operator: $T_2\rho \equiv
(I\otimes T)\rho \geq 0$. Here, $TB = B^T$ denotes the
transposition of the operator $B$ on ${\mathbb C}^{N_2}$ which is
defined in terms of the basis states of the second spin by means
of $\langle j_2,m_2 |B^T| j_2,m'_2\rangle=\langle j_2,m'_2 |B|
j_2,m_2\rangle$. The partial transposition $T_2$ is then defined
by $T_2(A\otimes B)=A\otimes B^T$.

The operation of taking the partial transposition destroys the
rotational invariance of states, i.~e., if $\rho$ is invariant
under rotations the partially transposed state $T_2\rho$ is
generally not $\mathrm{SO(3)}$-invariant. However, there exist
another operation which is unitarily equivalent to $T_2$ and which
does map rotationally invariant operators to rotationally
invariant operators. This operation will be denoted by
$\vartheta_2 = I\otimes \vartheta$. It involves the antiunitary
time reversal transformation $\vartheta$ of the second spin and
will therefore be referred to as partial time reversal.

According to Wigner's representation theorem \cite{WIGNER} the
action of the time reversal transformation $\vartheta$ on an
operator $B$ can be expressed as:
\begin{equation} \label{DEF-VARTHETA}
 \vartheta B = V B^T V^{\dagger} = \tau B^{\dagger} \tau^{-1}.
\end{equation}
In the first expression $T$ denotes again the transposition and
$V$ is a unitary matrix which represents a rotation of the
coordinate system about the $y$-axis by the angle $\pi$. In the
second expression of Eq.~(\ref{DEF-VARTHETA}) $\tau$ denotes the
operator $\tau=V\tau_0$ which is composed of the $\pi$-rotation
$V$ and of the operator $\tau_0$ of the complex conjugation. The
operator $\tau$ is antiunitary and satisfies
\begin{equation} \label{TAU-SQUARED}
 \tau^2 = (-1)^{2j_2}.
\end{equation}
$\vartheta$ is a positive but not completely positive map. It is
unitarily equivalent to the transposition $T$ and, hence, the
Peres-Horodecki criterion can be expressed by
\begin{equation} \label{PPT-CRITERION}
 \vartheta_2 \rho \equiv (I \otimes \vartheta) \rho \geq 0.
\end{equation}

A great advantage of the representation of states in $\beta$-space
is that the operators $Q_K$ have a very simple behaviour under the
map $\vartheta_2$. Namely, as is shown in Appendix \ref{APP-1}
they are eigenoperators of the partial time reversal:
$\vartheta_2Q_K = (-1)^{K}Q_K$. In $\beta$-space the map
$\vartheta_2$ therefore induces a reflection of the coordinate
axes corresponding to the odd values of $K$:
\begin{equation}
 \vartheta_2: \; \beta_K \mapsto (-1)^K \beta_K.
\end{equation}
We thus get the image $\vartheta_2S^{\beta}$ of $S^{\beta}$ by
reversing the signs of the odd coordinates.

We define $S^{\beta}_{\mathrm{ppt}}$ as the set of states which
are positive under $\vartheta_2$ or, equivalently, under $T_2$
(PPT states). This set is equal to the intersection of $S^{\beta}$
with its image $\vartheta_2S^{\beta}$. According to the
Peres-Horodecki criterion the set of separable states is a subset
of the set of PPT states. Hence, we have
\begin{equation}
 S^{\beta}_{\mathrm{sep}} \subset S^{\beta}_{\mathrm{ppt}}
 = S^{\beta} \cap \vartheta_2S^{\beta}.
\end{equation}

We note three properties which turn out to be useful in the
construction of the set of separable states.

{\textbf{(1)}} The functions defined by
Eq.~(\ref{DEF-FUNC-BETA-K}) are invariant under simultaneous
rotations of the input arguments:
\begin{equation} \label{DEF-INVAR}
 \tilde{\beta}_K[D^{(j_1)}(R)\varphi^{(1)},D^{(j_2)}(R)\varphi^{(2)}]
 = \tilde{\beta}_K[\varphi^{(1)},\varphi^{(2)}].
\end{equation}
This property is an immediate consequence of the rotational
invariance of the operators $Q_K$.

{\textbf{(2)}} The range $W^{\beta}$ defined in
Eq.~(\ref{DEF-W-BETA}) is obviously invariant under the partial
time reversal $\vartheta_2$. This means that $\bm{\beta}\in
W^{\beta}$ implies $\vartheta_2\bm{\beta}\in W^{\beta}$.

{\textbf{(3)}} There exist two distinguished separable states.
These are the state given by the parameter vector $\bm{\alpha}$
with components
\begin{equation} \label{DEF-ALPHA-JMAX}
 \alpha_J = \sqrt{\frac{N_1N_2}{2J_{\max}+1}} \delta_{J,J_{\max}},
 \qquad J_{\max} \equiv j_1+j_2,
\end{equation}
and the partially time reversed state given by
$\bm{\alpha}'=\vartheta_2\bm{\alpha}$. To proof this statement we
consider a pure product state $\rho$ of the form of
Eq.~(\ref{PROD-STATE}) with
$|\varphi^{(1)}\rangle=|j_1,+j_1\rangle$ and
$|\varphi^{(2)}\rangle=|j_2,+j_2\rangle$. We then have the obvious
relation
$|J=J_{\max},M=+J_{\max}\rangle=|\varphi^{(1)}\varphi^{(2)}\rangle$
and, hence,
\begin{equation}
 \langle\varphi^{(1)}\varphi^{(2)}|P_J|\varphi^{(1)}\varphi^{(2)}\rangle
 = \delta_{J,J_{\max}}.
\end{equation}
Equation (\ref{ALPHA-SEP}) then immediately leads to
Eq.~(\ref{DEF-ALPHA-JMAX}). This means that the pure product state
$\rho$ is mapped under the twirl operation to the separable state
$\Pi\rho = \frac{1}{2J_{\max}+1}P_{J_{\max}}$ corresponding to the
maximal value of the total angular momentum $J_{\max}$. It follows
from point {\textbf{(2)}} that also the partially time reversed
state is separable.

The point $\bm{\alpha}$ given by Eq.~(\ref{DEF-ALPHA-JMAX}) is an
extreme point of the simplex $S^{\alpha}$ and its image
$\bm{\alpha}'$ is an extreme point of $\vartheta_2S^{\alpha}$.
Thus, $\bm{\alpha}$ and $\bm{\alpha}'$ are extreme points of
$S^{\alpha}_{\mathrm{ppt}}$. It follows that the corresponding
points $\bm{\beta}=L\bm{\alpha}$ and $\bm{\beta}'=L\bm{\alpha}'$
in $\beta$-space belong to $W^{\beta}$ and represent extreme
points of $S^{\beta}_{\mathrm{ppt}}$.

As an illustration of the above analysis consider a $2\otimes N_2$
system for which $j_1=\frac{1}{2}$ and $j_2$ is arbitrary. As has
been demonstrated by Schliemann \cite{SCHLIEMANN1} the PPT
criterion is a necessary and sufficient separability condition in
this case. Within the present formulation this statement can be
proven as follows. We first note that the index $K$ takes on the
two values $K=0,1$ such that $\bm{\beta}$ is a two-dimensional
vector. Because of the normalization condition (\ref{NORM-BETA})
we only need a single parameter $\beta_1$ to characterize uniquely
an invariant state of a $2\otimes N_2$ system. It follows that
$S^{\beta}$ can be represented by an interval of the
$\beta_1$-axis, and $S^{\beta}_{\mathrm{ppt}}$ by a sub-interval
of this interval. Since an interval has exactly two extreme points
(its endpoints) we conclude with the help of point {\textbf{(3)}}
above that the extreme points of $S^{\beta}_{\mathrm{ppt}}$ belong
to $W^{\beta}$. By the relation (\ref{HULL}) the sets
$S^{\beta}_{\mathrm{ppt}}$ and $S^{\beta}_{\mathrm{sep}}$
therefore coincide. This shows that the PPT criterion is indeed
necessary and sufficient for separability.

\section{$3 \otimes N$ systems}\label{3tensorN}
Let us now consider the case $j_1=1$ ($N_1=3$) and $j_2$
arbitrary, i.~e. the case of $3 \otimes N_2$ systems. For
convenience we write $N\equiv N_2 = 2j_2+1$. Since $J$ takes on
the values $J=j_2-1$, $j_2$ and $j_2+1$, $\bm{\alpha}$ is a
three-vector
\begin{equation}
 \bm{\alpha} =
 \left( \begin{array}{c}
 \alpha_{j_2-1} \\ \alpha_{j_2} \\ \alpha_{j_2+1} \end{array} \right).
\end{equation}
The set $S^{\alpha}$ of invariant states is given by the
relations:
\begin{equation}
 \alpha_{j_2-1}, \alpha_{j_2}, \alpha_{j_2+1} \geq 0
\end{equation}
and
\begin{equation}
 \sqrt{\frac{N-2}{3N}}\alpha_{j_2-1}
 +\sqrt{\frac{1}{3}}\alpha_{j_2}+\sqrt{\frac{N+2}{3N}}\alpha_{j_2+1} = 1.
\end{equation}
We observe that $S^{\alpha}$ is a 2-simplex, i.~e. a triangle
whose vertices are given by the following parameter vectors
$\bm{\alpha}$:
\begin{equation} \label{VERTEX-S-A}
 \left( \begin{array}{c}
 0 \\ 0 \\ \sqrt{\frac{3N}{N+2}} \end{array} \right), \qquad
 \left( \begin{array}{c}
 \sqrt{\frac{3N}{N-2}} \\ 0 \\ 0 \end{array} \right), \qquad
 \left( \begin{array}{c}
 0 \\ \sqrt{3} \\ 0 \end{array} \right).
\end{equation}

In order to transform to $\beta$-space we first determine the
matrix $L$ by means of the formulae (\ref{L-0J})-(\ref{L-2J}):
\begin{eqnarray*} \label{L-3N}
 \lefteqn{ L = } \\
 && \left[\begin{array}{ccc}
  \sqrt{\frac{N-2}{3N}} & \sqrt{\frac{1}{3}}& \sqrt{\frac{N+2}{3N}} \\
 -\sqrt{\frac{(N-2)(N+1)}{2N(N-1)}} & -\sqrt{\frac{2}{(N-1)(N+1)}} &
 \sqrt{\frac{(N-1)(N+2)}{2N(N+1)}} \\
 \sqrt{\frac{(N+1)(N+2)}{6N(N-1)}} & -\sqrt{\frac{2(N-2)(N+2)}{3(N-1)(N+1)}} &
 \sqrt{\frac{(N-1)(N-2)}{6N(N+1)}}
 \end{array}\right].
\end{eqnarray*}
The extreme points of $S^{\beta}$ are found by applying this
matrix to the vectors given in Eq.~(\ref{VERTEX-S-A}). Since
$\beta_0$ is identically equal to $1$ by the normalization
condition (\ref{NORM-BETA}) we can represent points in
$\beta$-space by two coordinates $(\beta_1,\beta_2)$. One finds
that $S^{\beta}$ is a triangle in the $(\beta_1,\beta_2)$-plane
with the vertices:
\begin{equation} \label{DEF-A}
 A =
 \left(\sqrt{\frac{3(N-1)}{2(N+1)}},\sqrt{\frac{(N-1)(N-2)}{2(N+1)(N+2)}}\right),
\end{equation}
\begin{equation} \label{DEF-B}
 B =
 \left(-\sqrt{\frac{3(N+1)}{2(N-1)}},\sqrt{\frac{(N+1)(N+2)}{2(N-1)(N-2)}}\right),
\end{equation}
\begin{equation} \label{DEF-C}
 C =
 \left(-\sqrt{\frac{6}{(N-1)(N+1)}},-\sqrt{\frac{2(N-2)(N+2)}{(N-1)(N+1)}}\right).
\end{equation}

\begin{figure}[htb]
\includegraphics[width=\linewidth]{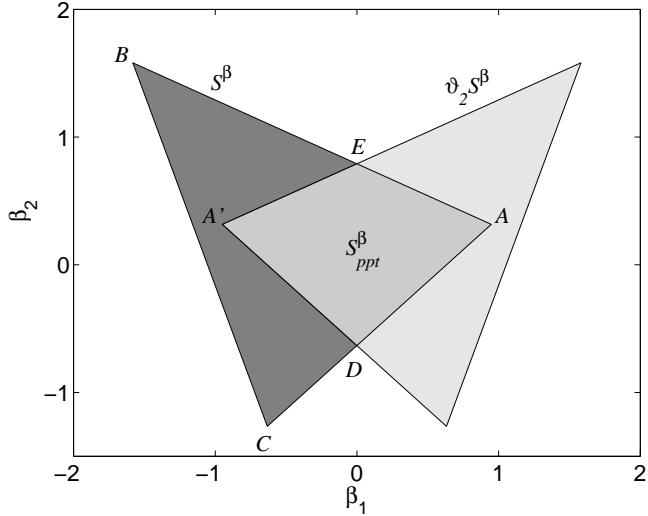}
\caption{State space structure of a system composed of two
particles with spins $j_1=1$ and $j_2=\frac{3}{2}$ ($N=4$). The
triangle $ABC$ represents the set $S^{\beta}$ of invariant states,
while the triangle $\vartheta_2S^{\beta}$ is its image under the
partial time reversal. The polygon $AA'DE$ represents the set
$S^{\beta}_{\mathrm{ppt}}$ of the PPT states. \label{fig1}}
\end{figure}

\begin{figure}[htb]
\includegraphics[width=\linewidth]{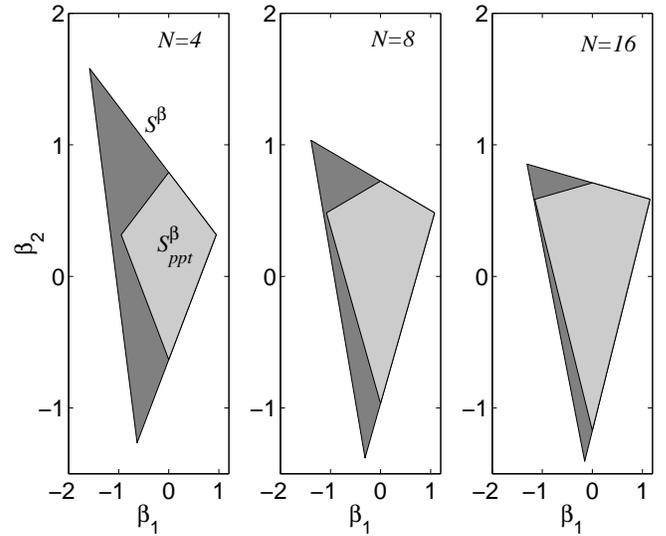}
\caption{The sets of the invariant states $S^{\beta}$ and of the
invariant PPT states $S^{\beta}_{\mathrm{ppt}}$ for three
different values of $N$. \label{fig2}}
\end{figure}

The image $\vartheta_2S^{\beta}$ of $S^{\beta}$ under the partial
time reversal is obtained by reversing the sign of the coordinate
$\beta_1$. Consequently, $S^{\beta}_{\mathrm{ppt}}$ is a polygon
with the four vertices $A$, $A'$, $D$ and $E$, where $A$ is given
by Eq.~(\ref{DEF-A}) and:
\begin{eqnarray}
 A'&=&\left(-\sqrt{\frac{3(N-1)}{2(N+1)}},
 \sqrt{\frac{(N-1)(N-2)}{2(N+1)(N+2)}}\right),
 \label{DEF-AP} \\
 D&=&\left(0,-\sqrt{\frac{2(N-1)(N-2)}{(N+1)(N+2)}}\right),
 \label{DEF-D} \\
 E&=&\left(0,\sqrt{\frac{(N+1)(N-1)}{2(N+2)(N-2)}}\right).
 \label{DEF-E}
\end{eqnarray}
Here, $A'=\vartheta_2A$ is the image of $A$ under $\vartheta_2$,
while $D$ and $E$ are the intersections of the lines $AC$ and $AB$
with the $\beta_2$-axis, respectively. The case $N=4$ is
illustrated in Fig.~\ref{fig1}. Similar pictures are obtained for
other values of $N$. Examples are shown in Fig.~\ref{fig2}. Note
that the origin of the $(\beta_1,\beta_2)$-plane describes the
state $\rho=\frac{1}{3N}I\otimes I$ of maximal entropy.

To construct the set $S^{\beta}_{\mathrm{sep}}$ of separable
states we have to investigate the functions:
\begin{eqnarray} \label{BETA-FUNC-1}
 \lefteqn{ \tilde{\beta}_1[\varphi^{(1)},\varphi^{(2)}] } \\
 && =\sqrt{N} \sum_{q=-1}^{+1}
 \langle\varphi^{(1)}|T^{(1)}_{1q}|\varphi^{(1)}\rangle
 \langle\varphi^{(2)}|T^{(2)\dagger}_{1q}|\varphi^{(2)}\rangle
 \nonumber
\end{eqnarray}
and
\begin{eqnarray} \label{BETA-FUNC-2}
 \lefteqn{ \tilde{\beta}_2[\varphi^{(1)},\varphi^{(2)}] } \\
 && =\sqrt{\frac{3N}{5}} \sum_{q=-2}^{+2}
 \langle\varphi^{(1)}|T^{(1)}_{2q}|\varphi^{(1)}\rangle
 \langle\varphi^{(2)}|T^{(2)\dagger}_{2q}|\varphi^{(2)}\rangle.
 \nonumber
\end{eqnarray}
We distinguish two cases, namely the cases of odd and of even $N$.
\newtheorem{theo}{Theorem}
\begin{theo}
For integer spins $j_2=1,2,3,\ldots$ one has
$S^{\beta}_{\mathrm{ppt}}=S^{\beta}_{\mathrm{sep}}$. Hence, for
all $3 \otimes N$ systems with odd $N$ the PPT criterion
represents a necessary and sufficient condition for the
separability of rotationally invariant states.
\end{theo}
To proof this theorem we show that the vertices $A$, $A'$, $D$ and
$E$ of the polygon $S^{\beta}_{\mathrm{ppt}}$ belong to
$W^{\beta}$. The statement
$S^{\beta}_{\mathrm{ppt}}=S^{\beta}_{\mathrm{sep}}$ then follows
immediately from Eq.~(\ref{HULL}).

The point $A$ corresponds to the parameter vector $\bm{\alpha}$
given by Eq.~(\ref{DEF-ALPHA-JMAX}). It follows that this point as
well as the point $A'=\vartheta_2A$ belong to $W^{\beta}$. Hence,
it suffices to verify that $D,E \in W^{\beta}$.

To show that $E \in W^{\beta}$ we choose the states
\begin{equation} \label{STATES-E}
 |\varphi^{(1)}\rangle = |1,m_1=0\rangle, \qquad
 |\varphi^{(2)}\rangle = |j_2,m_2=0\rangle.
\end{equation}
According to the selection rules for the matrix elements of the
tensor operators (\ref{DEF-TKQ}) and to Eq.~(\ref{T10}) we have
that $\langle\varphi^{(1)}|T^{(1)}_{1q}|\varphi^{(1)}\rangle=0$
for $q=0,\pm 1$ and, therefore,
\begin{equation} \label{BETA-TILDE-1}
 \tilde{\beta}_1 = 0.
\end{equation}
On the other hand, the non-vanishing matrix elements of the
second-rank tensors are given by [see Eq.~(\ref{T20})]:
\begin{equation} \label{MAT-1}
 \langle\varphi^{(1)}|T^{(1)}_{20}|\varphi^{(1)}\rangle
 = -\frac{2}{\sqrt{6}},
\end{equation}
and
\begin{equation} \label{MAT-2}
 \langle\varphi^{(2)}|T^{(2)}_{20}|\varphi^{(2)}\rangle
 = \frac{-2\sqrt{5}j_2(j_2+1)}{\sqrt{(N+2)(N+1)N(N-1)(N-2)}},
\end{equation}
which yields:
\begin{eqnarray} \label{BETA-TILDE-2}
 \tilde{\beta}_2 &=& \sqrt{\frac{3N}{5}}
 \langle\varphi^{(1)}|T^{(1)}_{20}|\varphi^{(1)}\rangle
 \langle\varphi^{(2)}|T^{(2)}_{20}|\varphi^{(2)}\rangle \nonumber \\
 &=& \sqrt{\frac{(N+1)(N-1)}{2(N+2)(N-2)}}.
\end{eqnarray}
We see from Eqs.~(\ref{BETA-TILDE-1}), (\ref{BETA-TILDE-2}) and
(\ref{DEF-E}) that $(\tilde{\beta}_1,\tilde{\beta}_2)=E$ and,
hence, that the point $E$ belongs to $W^{\beta}$.

To show that also $D$ belongs to $W^{\beta}$ we take the states
\begin{equation} \label{STATES-D}
 |\varphi^{(1)}\rangle = |1,0\rangle, \qquad
 |\varphi^{(2)}\rangle = |j_2,+j_2\rangle.
\end{equation}
Since the state $|\varphi^{(1)}\rangle$ is the same as before,
Eqs.~(\ref{BETA-TILDE-1}) and (\ref{MAT-1}) hold true. Instead of
Eq.~(\ref{MAT-2}), however, we get
\begin{equation}
 \langle\varphi^{(2)}|T^{(2)}_{20}|\varphi^{(2)}\rangle =
 \frac{2\sqrt{5}[3j_2^2-j_2(j_2+1)]}{\sqrt{(N+2)(N+1)N(N-1)(N-2)}}.
\end{equation}
This gives
\begin{eqnarray}
 \tilde{\beta}_2 &=& \sqrt{\frac{3N}{5}}
 \langle\varphi^{(1)}|T^{(1)}_{20}|\varphi^{(1)}\rangle
 \langle\varphi^{(2)}|T^{(2)}_{20}|\varphi^{(2)}\rangle \nonumber \\
 &=& -\sqrt{\frac{2(N-1)(N-2)}{(N+1)(N+2)}}.
\end{eqnarray}
A comparison with Eq.~(\ref{DEF-D}) shows that
$(\tilde{\beta}_1,\tilde{\beta}_2) = D \in W^{\beta}$. This
concludes the proof of the theorem.

\begin{figure}[htb]
\includegraphics[width=\linewidth]{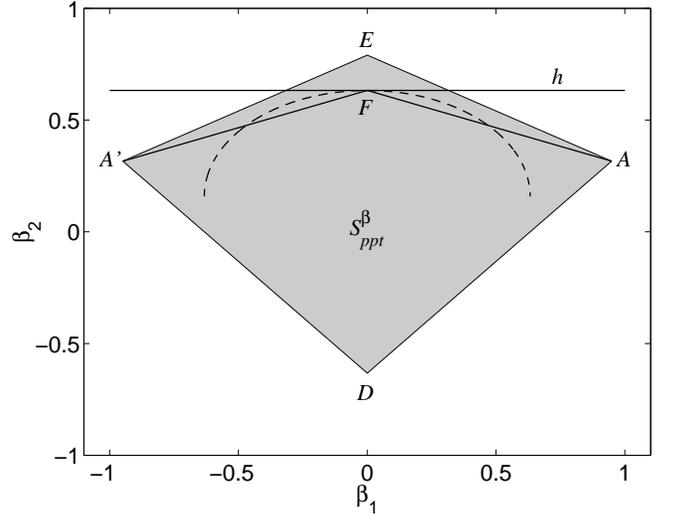}
\caption{The set of PPT states $S^{\beta}_{\mathrm{ppt}}$ for
$N=4$. The set $S^{\beta}_{\mathrm{sep}}$ lies entirely below the
straight line $h$ through $F$ which is parallel to the
$\beta_1$-axis. The broken line shows the curve defined by
Eqs.~(\ref{ELLIPSE-1}) and (\ref{ELLIPSE-2}). \label{fig3}}
\end{figure}

Let us now turn to the case of half-integer spins $j_2$, i.~e., we
assume that $N$ is even. Of course, we again have that $A$ and
$A'$ belong to $W^{\beta}$. But also $D \in W^{\beta}$ because the
state $|j_2,+j_2\rangle$ exists for integer as well as for
half-integer spins $j_2$. The argument following
Eq.~(\ref{STATES-D}) can thus also be applied in the present case.
It follows that $S^{\beta}_{\mathrm{sep}}$ contains at least the
triangle $AA'D$ (see Fig.~\ref{fig3}).

On the other hand, the state $|j_2,m_2=0\rangle$ exists, of
course, only for integer spins $j_2$. Instead of (\ref{STATES-E})
we consider the states
\begin{equation} \label{STATES-F}
 |\varphi^{(1)}\rangle = |1,0\rangle, \qquad
 |\varphi^{(2)}\rangle = |j_2,+1/2\rangle,
\end{equation}
which lead to
\begin{equation}
 \tilde{\beta}_1 = 0, \qquad
 \tilde{\beta}_2 = \sqrt{\frac{(N+2)(N-2)}{2(N+1)(N-1)}}.
\end{equation}
This shows that the point
\begin{equation} \label{DEF-F}
 F = \left(0,\sqrt{\frac{(N+2)(N-2)}{2(N+1)(N-1)}}\right)
\end{equation}
belongs to $W^{\beta}$. Hence, $S^{\beta}_{\mathrm{sep}}$ contains
at least the polygon with the vertices $A$, $A'$, $D$ and $F$.

We introduce the straight line $h$ which intersects the point $F$
and which is parallel to the $\beta_1$-axis (see Fig.~\ref{fig3}).
We are going to demonstrate that $S^{\beta}_{\mathrm{sep}}$ lies
entirely below this line. The line $h$ is thus tangential to
$S^{\beta}_{\mathrm{sep}}$ and corresponds to an optimal
entanglement witness (see Sec.~\ref{CONCLU}). To show this we
employ the rotational invariance of the functions
$\tilde{\beta}_K$ [see Eq.~(\ref{DEF-INVAR})] to obtain a suitable
parametrization of the states of the first spin $j_1=1$. Namely,
by an appropriate rotation $R$ any state of this spin can be
brought into the following form:
\begin{equation} \label{PHI-REP}
 |\varphi^{(1)}\rangle = \sqrt{r}|1,+1\rangle +
 \sqrt{1-r}|1,-1\rangle,
\end{equation}
where we omit an irrelevant overall phase factor and $r$ is a real
parameter taken from the interval $[0,1]$. Invoking the rotational
invariance we may assume without restriction that
$|\varphi^{(1)}\rangle$ is of this form. The state space of the
first spin $j_1$ has thus only a single relevant parameter $r\in
[0,1]$.

By use of the representation (\ref{PHI-REP}) the quantities
$\tilde{\beta}_1$ and $\tilde{\beta}_2$ become functions of the
parameter $r$ and of the state vector $|\varphi^{(2)}\rangle$ of
the second spin. Inserting Eq.~(\ref{PHI-REP}) into
Eq.~(\ref{BETA-FUNC-1}) and using Eqs.~(\ref{T10}) and (\ref{T11})
of Appendix~\ref{APP-1} we get
\begin{equation} \label{BETA1}
 \tilde{\beta}_1[r,\varphi^{(2)}]
 = \sqrt{\frac{N}{2}}(2r-1)
 \langle\varphi^{(2)}|T^{(2)}_{10}|\varphi^{(2)}\rangle.
\end{equation}
The function $\tilde{\beta}_2$ is found by substituting the
expression (\ref{PHI-REP}) into Eq.~(\ref{BETA-FUNC-2}) and by
using Eqs.~(\ref{T20})-(\ref{T22}). One finds that
$\tilde{\beta}_2$ can be written as the expectation value
\begin{equation} \label{BETA-EXPEC}
 \tilde{\beta}_2[r,\varphi^{(2)}]
 = \langle\varphi^{(2)}|H(\lambda)|\varphi^{(2)}\rangle
\end{equation}
of the Hermitian $(N \times N)$ matrix
\begin{equation} \label{DEF-H-LAMBDA}
 H(\lambda) \equiv H_0 + \lambda H_1.
\end{equation}
Here, we have defined
\[
 H_0 = \sqrt{\frac{N}{10}}T^{(2)}_{20}, \qquad
 H_1 = \frac{1}{2}
 \sqrt{\frac{3N}{5}}\left(T^{(2)}_{22}+T^{(2)\dagger}_{22}\right),
\]
and introduced the parameter
\begin{equation}
 \lambda = 2\sqrt{r(1-r)}, \qquad 0 \leq \lambda \leq 1.
\end{equation}

\begin{figure}[htb]
\includegraphics[width=\linewidth]{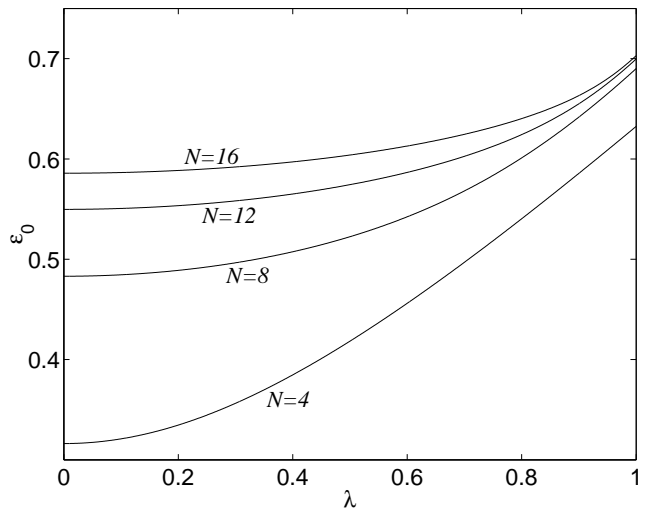}
\caption{The largest eigenvalue $\varepsilon_0(\lambda)$ of the
matrix $H(\lambda)$ defined by Eq.~(\ref{DEF-H-LAMBDA}) for
different values of $N$. \label{fig4}}
\end{figure}

For a given value of $\lambda$ the function $\tilde{\beta}_2$
defined by Eq.~(\ref{BETA-EXPEC}) is certainly smaller than or
equal to the largest eigenvalue of $H(\lambda)$ which we denote by
$\varepsilon_0(\lambda)$. We are going to demonstrate below that
$\varepsilon_0(\lambda)$ is a monotonically increasing function of
$\lambda$ and attains its maximum at $\lambda=1$:
\begin{equation} \label{E-MAX}
 \varepsilon_0(1) = \sqrt{\frac{(N+2)(N-2)}{2(N+1)(N-1)}}.
\end{equation}
Hence, we have
\begin{equation}
 \tilde{\beta}_2[r,\varphi^{(2)}] \leq \varepsilon_0(1)
\end{equation}
for all $r$ and $|\varphi^{(2)}\rangle$. Note that
$\varepsilon_0(1)$ is equal to the $\beta_2$-coordinate of the
point $F$ [see Eq.~(\ref{DEF-F})]. This shows that, as claimed,
all points of $W^{\beta}$ and, hence, all points of
$S^{\beta}_{\mathrm{sep}}$ lie below the line $h$.

To prove that $\varepsilon_0(\lambda)$ is a monotonically
increasing function of $\lambda$ we denote the eigenvalues of
$H(\lambda)$ by $\varepsilon_n(\lambda)$, where $n=0,1,2\ldots$,
and $n=0$ labels the largest eigenvalue. With the help of
Eq.~(\ref{THETA-T}) one verifies that $H(\lambda)$ is invariant
under time reversal. It follows that if $|\varphi\rangle$ is an
eigenstate of $H(\lambda)$ then also the time reversed state
$\tau|\varphi\rangle$ is an eigenstate with the same eigenvalue.
Since $j_2$ is half-integer valued the states $|\varphi\rangle$
and $\tau|\varphi\rangle$ are orthogonal. In fact, using the
antiunitarity of $\tau$ and Eq.~(\ref{TAU-SQUARED}) we get
\[
 \langle\tau\varphi|\varphi\rangle
 = \langle\tau^2\varphi|\tau\varphi\rangle^{\ast}
 = (-1)^{2j_2} \langle \tau\varphi|\varphi\rangle
 = -\langle\tau\varphi|\varphi\rangle,
\]
which shows that $\langle\tau\varphi|\varphi\rangle=0$.

All eigenvalues $\varepsilon_n(\lambda)$ are thus two-fold
degenerate and we write the corresponding eigenstates as
$|\varphi_{n,k}(\lambda)\rangle$, where the index $k=1,2$ labels
the eigenstates corresponding to the same eigenvalue:
$|\varphi_{n,2}(\lambda)\rangle =
\tau|\varphi_{n,1}(\lambda)\rangle$. We remark that the two-fold
degeneracy is analogous to the Kramers degeneracy according to
which the energy levels of an invariant system of an odd number of
spin-$\frac{1}{2}$ particles are at least two-fold degenerate
(see, e.~g.~\cite{SAKURAI}).

The Hellman-Feynman theorem now yields
\begin{equation} \label{E-DERIV1}
 \frac{d\varepsilon_0}{d\lambda} =
 \langle\varphi_{0,1}(\lambda)|H_1|\varphi_{0,1}(\lambda)\rangle.
\end{equation}
In particular, we have
\begin{equation} \label{E-DERIV1-0}
 \left.\frac{d\varepsilon_0}{d\lambda}\right|_{\lambda=0} = 0.
\end{equation}
On differentiating Eq.~(\ref{E-DERIV1}) once again we find:
\begin{equation}
 \frac{d^2\varepsilon_0}{d\lambda^2} = 2\sum_{n\neq 0,k}
 \frac{|\langle\varphi_{n,k}(\lambda)|H_1|\varphi_{0,1}(\lambda)\rangle|^2}
 {\varepsilon_0(\lambda)-\varepsilon_n(\lambda)} \geq 0.
\end{equation}
This shows that $\varepsilon_0(\lambda)$ is a convex function of
$\lambda$ with zero derivative at $\lambda=0$. It follows that
$\varepsilon_0(\lambda)$ increases monotonically. Some examples of
the behaviour of this functions are shown in Fig.~\ref{fig4}.

It remains to verify Eq.~(\ref{E-MAX}). We first note that $H(1)$
can be written with the help of Eqs.~(\ref{T20}) and (\ref{T22})
in terms of the spin operator $\hat{\bm{j}}^{(2)}$ as:
\begin{eqnarray} \label{H-1}
 H(1) &=& 2\sqrt{\frac{2}{(N+2)(N+1)(N-1)(N-2)}} \nonumber \\
 && \times \left( \left[\hat{\bm{j}}^{(2)}\right]^2
 - 3 \left[\hat{j}_y^{(2)}\right]^2 \right).
\end{eqnarray}
The largest eigenvalue of this matrix is given by
\begin{eqnarray} \label{E-MAX-1/2}
 \varepsilon_0(1)
 &=& 2\sqrt{\frac{2}{(N+2)(N+1)(N-1)(N-2)}} \nonumber \\
 && \times \left(j_2(j_2+1)-\frac{3}{4}\right).
\end{eqnarray}
Using $N=2j_2+1$ one shows that this equation coincides with
Eq.~(\ref{E-MAX}).

We finally demonstrate that the boundary of
$S^{\beta}_{\mathrm{sep}}$ is differentiable at the point $F$ [see
Eq.~(\ref{DEF-F})]. To this end, we construct a smooth curve which
belongs to $W^{\beta}$ and passes the point $F$. Consider the
following fixed state of the second spin:
\begin{equation}
 |\varphi^{(2)}\rangle = \frac{1}{\sqrt{2}}
 \left|\hat{j}^{(2)}_y=+1/2\right\rangle
 +\frac{i}{\sqrt{2}}\left|\hat{j}^{(2)}_y=-1/2\right\rangle.
\end{equation}
This is an eigenstate of the matrix $H(1)$ [Eq.~(\ref{H-1})]
corresponding to the largest eigenvalue $\varepsilon_0(1)$. Since
$|\varphi^{(2)}\rangle$ is fixed the functions $\tilde{\beta}_1$
and $\tilde{\beta}_2$ depend only on the parameter $r$ and
describe a curve in the $(\beta_1,\beta_2)$-plane. Writing
$r\equiv (1+\mu)/2$ and determining the matrix elements one finds:
\begin{eqnarray}
 \tilde{\beta}_1
 &=& \sqrt{\frac{3N^2}{8(N+1)(N-1)}}\mu, \label{ELLIPSE-1} \\
 \tilde{\beta}_2
 &=& \frac{\varepsilon_0(1)}{4}
 \left(1+3\sqrt{1-\mu^2}\right), \label{ELLIPSE-2}
\end{eqnarray}
where $-1\leq\mu\leq +1$. The curve described by these equations
represents the upper half of an ellipse in the
$(\beta_1,\beta_2)$-plane (see Fig.~\ref{fig3}). It intersects the
point $F$ and lies entirely in $W^{\beta}$. Since $F$ is the only
point of $h$ belonging to $W^{\beta}$, it follows that the
boundary of the separability region must be curved and that it is
differentiable at the extreme point $F$, the line $h$ being the
tangent. Summarizing, we have shown:
\begin{theo}
For half-integer spins
$j_2=\frac{3}{2},\frac{5}{2},\frac{7}{2},\ldots$ the set
$S^{\beta}_{\mathrm{sep}}$ of separable states is a true subset of
the set of PPT states. Hence, for all $3 \otimes N$ systems with
even $N$ the PPT criterion is only necessary and there always
exist entangled PPT states. The line $h$ represents the tangent to
$S^{\beta}_{\mathrm{sep}}$ at the extreme point $F$. The set
$S^{\beta}_{\mathrm{sep}}$ is bounded by the straight lines $AD$
and $A'D$ and by a concave curve which passes the points $A$, $A'$
and $F$.
\end{theo}

\section{Discussion and conclusions}\label{CONCLU}
The state space structure of rotationally invariant spin systems
has been analyzed in this paper. The set of invariant states has
been represented by means of two systems of invariant operators,
namely by the projections $P_J$ onto the total angular momentum
manifolds and by the invariant operators $Q_K$ composed of the
spherical tensors. The transformation between both representations
was found to be given by a matrix $L$ which is determined by
certain $6$-$j$ symbols of Wigner. The $Q_K$-representation is
particularly useful in applying the PPT criterion for separability
because the $Q_K$ are eigenoperators of the partial time reversal.
The method has been demonstrated to lead to a complete
classification of separability of $3 \otimes N$ systems. We have
shown that the PPT criterion is necessary and sufficient for all
system with odd $N$, while entangled PPT states exist for systems
with even $N$.

Some remarks on the structure of the state space in the limit $N
\rightarrow \infty$ might be of interest. In this limit the value
of the second spin $j_2$ becomes arbitrary large. We infer from
Eqs.~(\ref{DEF-B})-(\ref{DEF-D}) that the point $B$ then converges
to the point $A'$, and $C$ to $D$. At the same time $F$ converges
to $E$ [see Eqs.~(\ref{DEF-E}) and (\ref{DEF-F})]. Hence, as $N$
increases the set $S^{\beta}_{\mathrm{ppt}}$ approaches the set
$S^{\beta}$ and $S^{\beta}_{\mathrm{sep}}$ approaches
$S^{\beta}_{\mathrm{ppt}}$. This behaviour is also indicated in
Fig.~\ref{fig2}. The limit $N \rightarrow \infty$ thus corresponds
to a kind of classical limit in which all invariant states have a
positive partial transpose and are separable.

The line $h$ constructed in Sec.~\ref{3tensorN} leads to an
entanglement witness which we denote by ${\mathcal{W}}$. An
entanglement witness is a Hermitian operator which satisfies
${\mathrm{tr}}\{{\mathcal{W}}\sigma\}\geq 0$ for any separable
state $\sigma$, and ${\mathrm{tr}}\{{\mathcal{W}}\rho\}< 0$ for at
least one non-separable state $\rho$ \cite{HORODECKI96a,TERHAL}.
The hyperplane $h$ corresponding to an entanglement witness
${\mathcal{W}}$ is defined by
${\mathrm{tr}}\{{\mathcal{W}}\rho\}=0$. In the case of $3 \otimes
N$ systems $h$ is a one-dimensional line and the witness is, in
fact, optimal \cite{LEWENSTEIN00} because $h$ is tangential to the
region of separable states. We have formulated the witness in
$\beta$-space. Transforming back to $\alpha$-space one easily
shows that the entanglement witness corresponding to $h$ may be
written in terms of the projections $P_J$ as:
\begin{equation} \label{WITNESS}
 {\mathcal{W}} =
 -\frac{1}{N-2}P_{j_2-1}+P_{j_2}+\frac{1}{N+2}P_{j_2+1}.
\end{equation}
This expression leads to the following physical interpretation of
${\mathcal{W}}$. Suppose one carries out a measurement of the
total angular momentum $J$ on some invariant state $\rho$. If
$\rho$ is separable the inequality
\begin{equation} \label{INEQ}
 -\frac{p_{j_2-1}}{N-2}+p_{j_2}+\frac{p_{j_2+1}}{N+2} \geq 0
\end{equation}
must be satisfied, where $p_J={\mathrm{tr}}\{P_J\rho\}$ denotes
the probability of finding the value $J$. In other words, if the
inequality (\ref{INEQ}) is violated the state $\rho$ must
necessarily be entangled.

We exploit the witness (\ref{WITNESS}) to design a prescription
for the detection of entangled PPT states in $3 \otimes N$ systems
with even $N$ (bound entanglement \cite{HORODECKI98}). A given
state $\rho$ is positive under partial transposition if and only
if the corresponding point $(\beta_1,\beta_2)$ lies below the line
through $A'$ and $E$, and above the line through $A'$ and $D$ (see
Fig.~\ref{fig1}). If we transform to $\alpha$-space this yields
the conditions
\begin{equation} \label{INEQ-1}
 -\frac{2p_{j_2-1}}{N-1}+\frac{(N^2-5)p_{j_2}}{(N+1)(N-1)}
 +\frac{2p_{j_2+1}}{N+1} \geq 0
\end{equation}
and
\begin{equation} \label{INEQ-2}
 \frac{2p_{j_2-1}}{(N-1)(N-2)}-\frac{2p_{j_2}}{N-1}
 +p_{j_2+1} \geq 0.
\end{equation}
These inequalities are equivalent to the PPT condition
(\ref{PPT-CRITERION}). Hence, entangled PPT states can be detected
in the following way: Suppose again that a total angular momentum
measurement is performed on some state $\rho$. If one finds that
the measurement outcomes, i.~e. the probabilities $p_J$, satisfy
the inequalities (\ref{INEQ-1}) and (\ref{INEQ-2}) {\textit{and
violate}} the inequality (\ref{INEQ}) then the state $\rho$ must
be an entangled PPT state.

The witness ${\mathcal{W}}$ defined in Eq.~(\ref{WITNESS}) does
not detect all entangled PPT states. As has been shown in
Sec.~\ref{3tensorN} a part of the boundary of the region of the
separable states is curved and, therefore, one needs an infinite
number of linear entanglement witnesses. The upper boundary of
$S^{\beta}_{\mathrm{sep}}$ can of course be described by means of
a suitable nonlinear equation. A possible way to derive the latter
is to construct the envelope of appropriate families of curves of
the type given by Eqs.~(\ref{ELLIPSE-1}) and (\ref{ELLIPSE-2}).

The considerations of Sec.~\ref{3tensorN} reveal that for $3
\otimes N_2$ systems half-integer spins are crucial for the
emergence of entangled PPT states. The entanglement structure of
systems involving half-integer spins is thus quite different from
those with integer spins. It seems that this is closely connected
to the fact that pure states which are invariant under time
reversal only exist for integer spins, while for half-integer
spins a given pure state is always orthogonal to its time reversed
state. A clear physical interpretation of this result and its
generalization to arbitrary $N_1\otimes N_2$ systems is of great
interest. The next step to further investigate this point could be
to study $4\otimes N_2$ systems, which is possible by the method
developed in this paper.

\begin{acknowledgments}
The author would like to thank J. Schliemann and F. Petruccione
for helpful comments and stimulating discussions.
\end{acknowledgments}

\appendix

\section{Spherical tensor operators}\label{APP-1}
We define here the irreducible spherical tensor operators $T_{Kq}$
acting on the state space ${\mathbb C}^{N}$ of a particle with
spin $j$, where $N=2j+1$, $K=0,1,\ldots,2j$, and $q=-K,\ldots,+K$.
The tensor operators $T^{(i)}_{K_iq_i}$ for $i=1,2$ used in the
main text are obtained by setting $j=j_1$ or $j=j_2$.

The spherical tensor operators $T_{Kq}$ represent a complete
system of operators on ${\mathbb C}^{N}$. This means that any
operator on the state space of the spin-$j$ particle may be
written as a unique linear combination of the $T_{Kq}$. Moreover,
the tensors are orthonormal with respect to the Hilbert-Schmidt
inner product:
\begin{equation} \label{T-ORTHO}
 {\mathrm{tr}}\left\{T^{\dagger}_{K'q'}T_{Kq}\right\}
 = \delta_{KK'}\delta_{qq'}.
\end{equation}
For a fixed $K$ the $(2K+1)$ operators $T_{Kq}$ represent the
spherical components of a tensor of rank $K$. They transform
according to an irreducible representation of $\mathrm{SO(3)}$
which corresponds to the angular momentum $K$:
\begin{equation} \label{T-TRAFO}
 D^{(j)}(R) T_{Kq} D^{(j)}(R)^{\dagger} = \sum_{q'=-K}^{+K}
 D^{(K)}_{q'q}(R) T_{Kq'}.
\end{equation}
For instance, the $T_{1q}$ behave as components of a vector, and
the $T_{2q}$ as components of a second-rank tensor.

The matrix elements of the tensors may be defined in term of
Wigner's $3$-$j$ symbols as \cite{WIGNER,EDMONDS}
\begin{equation} \label{DEF-TKQ}
 \langle j,m|T_{Kq}|j,m'\rangle = \sqrt{2K+1} (-1)^{j-m}
 \left(\begin{array}{ccc}
 j & j & K \\
 m & -m' & -q
 \end{array}\right).
\end{equation}
The $3$-$j$ symbols are closely related to the Clebsch-Gordan
coefficients:
\begin{eqnarray} \label{DEF-3J}
 \lefteqn{ \langle j_1,m_1;j_2,m_2|JM\rangle = } \nonumber \\
 && \sqrt{2J+1} (-1)^{j_1-j_2+M}
 \left(\begin{array}{ccc}
 j_1 & j_2 & J \\
 m_1 & m_2 & -M
 \end{array}\right).
\end{eqnarray}
According to the selection rules for the $3$-$j$ symbols the
matrix element (\ref{DEF-TKQ}) is equal to zero for $m-m'-q \neq
0$. In particular, we have $T_{00}=\frac{1}{\sqrt{N}}I$.

The matrix elements (\ref{DEF-TKQ}) of the tensor operators are
real and one has $T^{\dagger}_{Kq} = T^T_{Kq} = (-1)^q T_{K,-q}$.
It follows that the $T_{Kq}$ are eigenoperators of the time
reversal transformation $\vartheta$ which was defined in
Eq.~(\ref{DEF-VARTHETA}). In fact, using the transformation
behaviour (\ref{T-TRAFO}) of the tensors and the fact that a
rotation by $\pi$ about the $y$-axis is represented by the unitary
matrix
\begin{equation} \label{ELEMENTS-V}
 D^{(K)}_{q'q}(\pi) = (-1)^{K-q'}\delta_{q',-q},
\end{equation}
one finds
\begin{equation} \label{THETA-T}
 \vartheta T_{Kq} = V T^T_{Kq} V^{\dagger}
 = (-1)^K T_{Kq}.
\end{equation}
As a consequence the operators $Q_K$ which have been introduced in
Eq.~(\ref{DEF-QK}) are eigenoperators of the partial time reversal
$\vartheta_2=I\otimes\vartheta$:
\begin{equation}
 \vartheta_2 Q_K = (-1)^K Q_K.
\end{equation}

We finally list the non-vanishing matrix elements of the tensor
operators needed in Sec.~\ref{3tensorN}:
\begin{equation} \label{T10}
 \langle j,m|T_{10}|j,m\rangle
 = 2m\sqrt{\frac{3}{N(N-1)(N+1)}},
\end{equation}
\begin{equation} \label{T11}
 \langle j,m|T_{11}^{\dagger}|j,m+1\rangle
 = -\sqrt{\frac{6(j-m)(j+m+1)}{N(N-1)(N+1)}},
\end{equation}
\begin{equation} \label{T20}
 \langle j,m|T_{20}|j,m\rangle
 = \frac{2\sqrt{5}[3m^2-j(j+1)]}{\sqrt{(N+2)(N+1)N(N-1)(N-2)}},
\end{equation}
\begin{eqnarray} \label{T21}
 \lefteqn{ \langle j,m|T_{21}^{\dagger}|j,m+1\rangle = } \\
 &&
 -\sqrt{5}(1+2m)\sqrt{\frac{6(j-m)(j+m+1)}{(N+2)(N+1)N(N-1)(N-2)}},
 \nonumber
\end{eqnarray}
\begin{eqnarray} \label{T22}
 \lefteqn{ \langle j,m|T_{22}^{\dagger}|j,m+2\rangle = } \\
 &&
 \sqrt{5}\sqrt{\frac{6(j-m-1)(j-m)(j+m+1)(j+m+2)}{(N+2)(N+1)N(N-1)(N-2)}}.
 \nonumber
\end{eqnarray}

\section{Proof of relation (\ref{L-6j})}\label{APP-2}
The starting point is given by Eq.~(\ref{L-REP}). We insert into
this equation the definitions (\ref{DEF-PJ}) and (\ref{DEF-QK})
for the invariant operators $P_J$ and $Q_K$, and introduce
complete sets of product basis states $|j_1,m_1;j_2,m_2\rangle$.
This yields a multiple sum over products of two Clebsch-Gordan
coefficients and two matrix elements of the tensor operators. By
use of Eqs.~(\ref{DEF-TKQ}) and (\ref{DEF-3J}) the Clebsch-Gordan
coefficients as well as the matrix elements of the spherical
tensors can be written in terms of the $3$-$j$ symbols. We also
use the selection rules for the $3$-$j$ symbols and their symmetry
properties. This procedure leads to the following sum over
$4$-fold products of $3$-$j$ symbols:
\begin{eqnarray} \label{DERIV-2}
 L_{KJ} &=& \sqrt{(2K+1)(2J+1)} (-1)^{j_1+j_2+J} \times \\
 &~& \sum_{\{m_i\}} \chi(\{m_i\}) \times \nonumber  \\
 &~& \;\;\;\;\;\;
 \left(\begin{array}{ccc}
 j_1 & j_2 & J \\
 m_1 & m_2 & m_3
 \end{array}\right)
 \left(\begin{array}{ccc}
 j_1 & j_1 & K \\
 -m_1 & m_5 & -m_6
 \end{array}\right) \times \nonumber \\
 &~& \;\;\;\;\;\;
 \left(\begin{array}{ccc}
 j_2 & j_2 & K \\
 -m_4 & -m_2 & m_6
 \end{array}\right)
 \left(\begin{array}{ccc}
 j_2 & j_1 & J \\
 m_4 & -m_5 & -m_3
 \end{array}\right), \nonumber
\end{eqnarray}
where $\chi(\{m_i\})$ is a phase factor:
\begin{eqnarray*}
 \chi(\{m_i\}) &=&
 (-1)^{j_1+m_1}(-1)^{j_2+m_2}(-1)^{J+m_3} \times \nonumber \\
 &~&
 (-1)^{j_2+m_4}(-1)^{j_1+m_5}(-1)^{K+m_6}.
\end{eqnarray*}
The sum over the quantum numbers $m_1,\ldots,m_6$ in
Eq.~(\ref{DERIV-2}) exactly corresponds to a certain $6$-$j$
symbol of Wigner \cite{WIGNER}. A general $6$-$j$ symbol involves
six angular momenta and is written as
\begin{equation} \label{DEF-6j}
 \left\{\begin{array}{ccc}
 j_1 & j_2 & j_3 \\
 j_4 & j_5 & j_6
 \end{array}\right\}.
\end{equation}
The sum of Eq.~(\ref{DERIV-2}) is equal to the $6$-$j$ symbol
(\ref{DEF-6j}) with $j_3=J$, $j_4=j_2$, $j_5=j_1$ and $j_6=K$.
Hence, we see that Eq.~(\ref{DERIV-2}) reduces to
Eq.~(\ref{L-6j}). We remark that a similar technique has been used
in Ref.~\cite{NtensorN} in order to derive an expression for the
matrix which represents the partial time reversal $\vartheta_2$ in
the $P_J$-representation.

By use of the formulae for the $6$-$j$ symbols \cite{EDMONDS} we
find that the first three rows of $L$ are given by
\begin{equation} \label{L-0J}
 L_{0J} = \sqrt{\frac{2J+1}{N_1N_2}},
\end{equation}
and
\begin{widetext}
\begin{equation} \label{L-1J}
 L_{1J} = -2\sqrt{3(2J+1)}
 \frac{j_1(j_1+1)+j_2(j_2+1)-J(J+1)}{\sqrt{(N_1-1)N_1(N_1+1)(N_2-1)N_2(N_2+1)}},
\end{equation}
\begin{equation} \label{L-2J}
 L_{2J} = 2\sqrt{5(2J+1)}
 \frac{3X(X-1)-4j_1(j_1+1)j_2(j_2+1)}{\sqrt{(N_1-2)(N_1-1)N_1(N_1+1)(N_1+2)
 (N_2-2)(N_2-1)N_2(N_2+1)(N_2+2)}},
\end{equation}
\end{widetext}
where $X \equiv j_1(j_1+1)+j_2(j_2+1)-J(J+1)$.

\end{document}